\documentclass[twocolumn,prl,aps,showpacs]{revtex4}
\usepackage{graphicx}

\begin{document}

\title{Boundary Condition of Polyelectrolyte Adsorption}

\author{Chi-Ho Cheng}
\email{phcch@phys.sinica.edu.tw} \affiliation{Institute of
Physics, Academia Sinica, Taipei, Taiwan}

\date{\today}

\begin{abstract}
The modification of the boundary condition for polyelectrolyte
adsorption on charged surface with short-ranged interaction is
investigated under two regimes. For weakly charged Gaussian
polymer in which the short-ranged attraction dominates, the
boundary condition is the same as that of the neutral polymer
adsorption. For highly charged polymer (compressed state) in which
the electrostatic interaction dominates, the linear relationship
(electrostatic boundary condition) between the surface monomer
density and the surface charge density needs to be modified.
\end{abstract}


\maketitle


\section{I. introduction}

Polyelectrolyte adsorption on neutral (due to short-ranged
interaction) and charged (due to electrostatic interaction)
surface is still active and important in recent years
\cite{barrat, grosberg, netz}. Theoretical approach on solving the
continuum theory (Edwards equation and its derivatives)
\cite{edwards} on the adsorption problem requires a proper
boundary condition.

The boundary condition for a pure short-ranged attraction was
first given by de Gennes \cite{deGennes}. Later, the same boundary
condition was adopted for problems with both short-ranged and
electrostatic interaction between the polymer and the surface
\cite{chatellier, joanny, shafir, biesheuvel,wang}. The treatment
implicitly assumes that the short-ranged interaction between the
polymer and the surface dominates over the electrostatic ones.
However, it is still a question for the validity of this
assumption.

On the other hand, it was recently identified that the boundary
condition for a highly charged polymer adsorbed on the charged
surface is governed by the electrostatic boundary condition, and
it can simply be expressed in a linear form between the surface
monomer density and the surface charged density in the adsorption
regime (compressed state) \cite{cheng04, cheng05a}. With an extra
perturbed short-ranged interaction, although the interaction is
also dominated by the electrostatic one, it is still a puzzle
whether the form of the boundary condition remains unchanged or
its modification is needed.

In this paper, we are going to fill the above two gaps in the
literature. We show that for a weakly charged polymer adsorption
due to short-ranged attraction, an perturbed electrostatic
interaction in general does not modify the boundary condition. For
a highly charged polymer adsorption (compressed state) due to
electrostatic interaction, an perturbed short-ranged interaction
would induce a non-linear correction to the original boundary
condition expressed in a linear form between the surface mononer
density and the surface charge density.

\section{II. short-ranged attraction revised}

Before our main investigation, we first revise a Gaussian polymer
adsorbed on the surface with short-ranged attraction. Suppose the
short-ranged attraction between the monomers and the hard-wall
surface is modelled by the $\delta$-potential $-\gamma
\delta(z-b)$ located just above the hard-wall at $z=0$.

The continuum equation describing the density profile
$\rho(z)=\psi_0^2(z)$ is determined by the Edwards equation
\begin{eqnarray} \label{edwards}
\left(-\frac{a^2}{6}\frac{d^2}{dz^2} - \beta \gamma
\delta(z-b)\right)\psi_0(z) = \varepsilon_0 \psi_0(z)
\end{eqnarray}
where $a$ is the bond length, $\beta=1/(k_{\rm B}T)$, and
$\varepsilon_0$ is the ground state eigenvalue. The boundary
condition imposed is $\psi_0(0)=0$ and $\psi_0(+\infty)=0$.
Similar to the usual eigenproblem appearing in Quantum Mechanics
\cite{landau},
\begin{eqnarray} \label{psi1a}
\psi_0(z)=\left\{
\begin{array}{cc}
\sinh(z/d_0) , & \ 0\leq z\leq b \\
A \exp(-z/d_0) , & \ z\geq b
\end{array}\right.
\end{eqnarray}
up to a normalization constant. $d_0$ describes the length scale
of the diffusion layer of the adsorbed polymer. By fitting the
boundary condition at $z=b$, we have
\begin{eqnarray} \label{d0}
\frac{b}{d_0} \left(1+ \coth(\frac{b}{d_0})\right)
 = \frac{6\beta \gamma b}{a^2}
\end{eqnarray}
The binding energy (in unit of $k_{\rm B}T)$, or the eigenvalue
$\varepsilon_0=-a^2/6d_0^2$.

The idea suggested by de Gennes \cite{deGennes} to absorb the
$\delta$-potential into the surface (by taking sufficiently small
$b$) is to modify the boundary condition at the surface and to
match with the asymptotic behavior away from the surface by
identifying the same binding energy (eigenvalue). That is, we are
looking at the profile
\begin{eqnarray} \label{psi1b}
\psi_1(z) = A \exp(-z/d_1),  \ \ 0 < z < +\infty
\end{eqnarray}
in which it is the solution of the eigenproblem
\begin{eqnarray}
-\frac{a^2}{6}\frac{d^2}{dz^2}\psi_1(z) = \varepsilon_0 \psi_1(z)
\end{eqnarray}
with the boundary condition
\begin{eqnarray} \label{d0a}
\left.\frac{1}{\psi_1}\frac{d\psi_1}{dz}\right|_{z=0^+} &=&
-\frac{1}{d_1}
\\ \psi_1(+\infty) &=& 0  \label{d0b}
\end{eqnarray}
which is adopted on neutral polymer adsorption. The binding energy
(in unit of $k_{\rm B}T)$ $\varepsilon_0=-a^2/6d_1^2$. Hence
$d_1=d_0$. Notice that, the microscopic parameters $\gamma$ and
$b$ are now replaced by the macroscopic quantity $d_0$.

\section{III. short-ranged attraction with perturbed electrostatic interaction}

Suppose the weakly charged polymer can still keep its Gaussian
features when an perturbed local electrostatic interaction $V(z)$
from the charged surface is considered. In general the local
potential $V(z)=V_0$ at $z=0$, becomes linear at $z \gtrsim 0$,
and saturate to zero at large $z\gtrsim r_{\rm s}$ ($r_{\rm s}$ is
the Debye screening length). The Edwards equation is
\begin{eqnarray} \label{edwards2a}
\left(-\frac{a^2}{6}\frac{d^2}{dz^2} - \beta \gamma \delta(z-b) +
\beta V(z) \right)\psi_0(z) = \varepsilon_0 \psi_0(z)
\end{eqnarray}
with the boundary condition $\psi_0(0)=\psi_0(+\infty)=0$.
Following the same spirit in previous section, we absorb the
$\delta$-potential into the surface such that the eigenproblem
becomes
\begin{eqnarray} \label{edwards2b}
\left(-\frac{a^2}{6}\frac{d^2}{dz^2} + \beta V(z) \right)\psi_1(z)
= \varepsilon_0 \psi_1(z)
\end{eqnarray}
with the boundary condition same as Eqs.(\ref{d0a})-(\ref{d0b}).
The binding energy $\varepsilon_0$ in both Eqs.(\ref{edwards2a})
and (\ref{edwards2b}) can be estimated by the first-order
perturbation theory \cite{landau} to the solution in
Eqs.(\ref{psi1a}) and (\ref{psi1b}), respectively. In
Eq.(\ref{edwards2a}), its corresponding eigenvalue
\begin{eqnarray} \label{binding}
\varepsilon_0 &=& -\frac{a^2}{6 d_0^2} + \left( \int_0^b +
\int_b^{\infty}\right) dz \
 \psi_0^2(z) \beta V(z)  \nonumber \\
 &\simeq& -\frac{a^2}{6 d_0^2} + \int_b^{\infty} dz \
 \psi_0^2(z) \beta V(z)
\end{eqnarray}
at sufficiently small $b$. The eigenvalue in Eq.(\ref{edwards2b})
shares the same form
\begin{eqnarray} \label{binding2}
\varepsilon_0 = -\frac{a^2}{6 d_1^2} + \int_0^{\infty} dz \
 \psi_1^2(z) \beta V(z)
\end{eqnarray}
except $d_0$ is replaced by $d_1$. Hence, by identifying the same
eigenvalue in both Eqs.(\ref{binding})-(\ref{binding2}), we get
$d_1 = d_0$. Both the neutral and weakly charged Gaussian polymer
share the same boundary condition due to short-ranged attractive
surface. Notice that the discussion of the boundary condition was
also made by Joanny in which the coupling of the monomer density
to a further electrostatic equation of Poisson-Boltzmann type is
considered. The effective $d_1$ would then be different from $d_0$
\cite{joanny}.

In order to investigate the validity of the Gaussian feature, we
choose the local potential of the Debye-H\"{u}ckel form $V(z)=V_0
\exp(-z/r_{\rm s})$, where $V_0=4\pi l_{\rm B} \tau \sigma r_{\rm
s}$, with the Bjerrum length $l_{\rm B}$, line charge density of
polymer $\tau$, and surface charge density of the surface
$\sigma$. Substituting this $V(z)$ into Eq.(\ref{binding2}),
\begin{eqnarray}
\varepsilon_0 = -\frac{a^2}{6 d_0^2} + \frac{2\beta V_0 r_{\rm
s}}{2 r_{\rm s} + d_0}
\end{eqnarray}
The first term is the binding energy due to short-ranged
attraction whereas the second term the electrostatic interaction.
The condition for perturbed electrostatic interaction requires
\begin{eqnarray} \label{condition}
\frac{a^2}{6 d_0^2} \gg \frac{2\beta |V_0| r_{\rm s}}{2 r_{\rm s}
+ d_0}
\end{eqnarray}
where it becomes $|V_0| \ll k_{\rm B}T a^2/6d_0^2$ for low ionic
strength $r_{\rm s} \gg d_1$. For high ionic strength $r_{\rm s}
\ll d_0$, it requires $|V_0| \ll k_{\rm B}T a^2/12 r_{\rm s} d_0$.
Eq.(\ref{condition}) is a necessary condition to identify whether
the electrostatic interaction is still perturbatively small. If
the surface charge density becomes strong such that $|V_0|$ no
longer satisfy Eq.(\ref{condition}), the Gaussian polymer
undergoes conformational changes. The corresponding boundary
condition would deviate Eq.(\ref{d0a}) very much.

\section{IV. electrostatic boundary condition with perturbed short-ranged interaction}

In another regime that the polymer is highly charged such that the
adsorbed polymer is in a compressed state on the substrate, the
boundary condition is determined by the electrostatic boundary
condition across the dielectric \cite{cheng04, cheng05a}.
The continuum theory is described also by the Edwards equation
\begin{eqnarray} \label{edwards4a}
\left(-\frac{a^2}{2}\frac{d^2}{dz^2}  + \beta V(z)
\right)\psi_0(z) = \varepsilon_0 \psi_0(z)
\end{eqnarray}
where the coefficient of the entropic term is $-a^2/2$ instead of
$-a^2/6$ \cite{cheng05a}. The boundary condition imposed is
$\psi_0(0)=C_0$ and $\psi_0(+\infty)=0$. $C_0 \neq 0$ because the
electrostatic boundary condition for a compressed adsorbed
polyelectrolyte needs to be satisfied \cite{cheng04},
\begin{eqnarray} \label{electrostatic}
C_0^2 = -\frac{2K}{\epsilon'/\epsilon-1} \left(\sigma +
\frac{\epsilon'/\epsilon+1}{2}\sigma_{\rm p} \right)
\end{eqnarray}
where $\epsilon$ and $\epsilon'$ are the dielectric constant of
the medium and the substrate, respectively. $\sigma$ is the
surface charge density just above the substrate. $\sigma_{\rm p}$
is the polarization surface charge density induced by the polymer
only. It depends on $\epsilon'/\epsilon$ but not on $\sigma$. $K$
is the proportional constant depending only on
$\epsilon'/\epsilon$. Both $K$ and $\sigma_{\rm p}$ are model
dependent; in other words, they depend on the microscopic details
of the system. Similar to the diffusive layer thickness $d$
appearing in the previous sections, the microscopic details are
absorbed into these two marcroscopic quantities $K$ and
$\sigma_{\rm p}$.

In the following, with the perturbed short-ranged interaction
(attractive or repulsive) modelled by a $\delta$-potential located
just above the substrate, we are going to investigate how this
perturbed term is adsorbed into the boundary condition. That is,
we consider the Edwards equation
\begin{eqnarray} \label{edwards4b}
\left(-\frac{a^2}{2}\frac{d^2}{dz^2} + \beta V(z) \right)\psi_1(z)
= \varepsilon_1 \psi_1(z)
\end{eqnarray}
with the boundary condition $\psi_1(0)=C_1$ and
$\psi_1(+\infty)=0$. Notice that $\varepsilon_0$ in
Eq.(\ref{edwards4a}) (without $\delta$-potential) is not equal to
$\varepsilon_1$ in Eq.(\ref{edwards4b}) (with $\delta$-potential).
In fact, the binding energy $\varepsilon_1$ can be related to
$\varepsilon_0$ by perturbation theory \cite{landau} up to first
order, in which
\begin{eqnarray}
\varepsilon_1 &=& \varepsilon_0 + \int_0^\infty dz
 \psi_0^2(z)(-\beta \gamma \delta(z-b))  \nonumber \\
&=& \varepsilon_0 -\beta \gamma \psi_0^2(b) \nonumber \\
&\rightarrow&  \varepsilon_0 -\beta \gamma C_0^2  \label{e1}
\end{eqnarray}
for sufficiently small $b$. The change of the surface mononer
density due to the perturbed interaction can be further estimated
by applying the WKB approximation \cite{landau}. Near the surface,
we have
\begin{eqnarray} \label{psi1}
\psi_0(z) &=& \frac{A}{(\varepsilon_0-V(z))^{1/4}}
 \sin\left(\frac{\sqrt{2}}{a} \int_0^z dz \sqrt{\varepsilon_0-V(z)} + \alpha\right)   \nonumber  \\
 &\simeq& \frac{A}{(\varepsilon_0-V_0)^{1/4}}
 \sin(\frac{\sqrt{2(\varepsilon_0-V_0)}}{a} z + \alpha)
\end{eqnarray}
where $\alpha \neq 0$ related to
\begin{eqnarray} \label{c0}
C_0 = \frac{A}{(\varepsilon_0-V_0)^{1/4}}\sin\alpha
\end{eqnarray}
Notice that, in the usual case of Quantum Mechanics \cite{landau},
because of the hard-wall boundary condition $C_0=0$, $\alpha$ is
set to be zero.

Similarly, we can also write
\begin{eqnarray} \label{psi2}
\psi_1(z) &\simeq& \frac{A}{(\varepsilon_1-V_0)^{1/4}}
 \sin(\frac{\sqrt{2(\varepsilon_1-V_0)}}{a} z + \alpha)
\end{eqnarray}
where the coefficients $A$ and $\alpha$ are assumed unchanged.
Hence
\begin{eqnarray} \label{c1}
C_1 = \frac{A}{(\varepsilon_1-V_0)^{1/4}}\sin\alpha
\end{eqnarray}
From Eqs.(\ref{c0}) and (\ref{c1}), we got the relation
$(\varepsilon_0-V_0) C_0^4 = (\varepsilon_1-V_0) C_1^4$, and hence
\begin{eqnarray} \label{c0c1}
C_1 &\simeq& C_0 -
\frac{C_0}{4(\varepsilon_0-V_0)}(\varepsilon_1-\varepsilon_0)
\nonumber
\\ &=& C_0 + \frac{\beta\gamma}{4(\varepsilon_0-V_0)} C_0^3
\end{eqnarray}
by applying Eq.(\ref{e1}). Remind that $\varepsilon_0-V_0 > 0$.
Eq.(\ref{c0c1}) is consistent with our picture that short-ranged
attraction (repulsion), $\gamma > 0$ ($< 0$), increases
(decreases) the surface mononer density. The next higher order
correction for $C_1$ is $O(C_0^3)$ \cite{remarks}. The linear
relation between the surface mononer density and the surface
charge density is no longer valid after including the short-ranged
interaction effect. However, the violation of the linear relation
implies that part of the surface mononer density is not due to the
electrostatic interaction in which the electrostatic boundary
condition does not apply \cite{remarks2}.

\section{acknowledgements}
The author would like to thank P.Y. Lai and X. Wu for helpful
comments. The work was supported by the Postdoctoral Fellowship of
Academia Sinica.




\begin{thebibliography}{}
\bibitem{remarks}
The author have tried several different ``variational
wavefunctions" $\psi(z)$ to determine the next higher order
correction, and it was found that the orders are different among
those different variational wavefunctions. It seems that the order
depends on the details of the potential $V(z)$. However, the main
point is that the correction should be at higher order, and hence
the simple linear form of the boundary condition no longer holds
after including the short-ranged interaction.
\bibitem{remarks2}
Another way to understand it is as follows. The part of the
adsorbed monomers due to short-ranged interaction $-\gamma
\delta(z-b)$ is adsorbed at $z=b$, and it is counted into the
surface monomer density after taking sufficiently small $b$.
\end{thebibliography}
\end{document}